# Surface activation of Hastalex by vacuum argon plasma for cytocompatibility enhancement

**Nikola Slepičková Kasálková[a], Silvie Rimpelová[b], Cyril Vacek[a], Dominik Fajstavr[a], Václav Švorčík[a], Petr Sajdl[c], Petr Slepička[a,*]**

[a]Department of Solid State Engineering, University of Chemistry and Technology Prague, 166 28 Prague, Czech Republic

[b]Department of Biochemistry and Microbiology, University of Chemistry and Technology Prague, 166 28 Prague, Czech Republic

[c]Department of Power Engineering, University of Chemistry and Technology Prague, 166 28 Prague, Czech Republic

**Abstract**

Here, we present surface analysis and biocompatibility evaluation of novel composite material based on graphene oxide traded as Hastalex. First, the surface morphology and elemental analysis of the pristine material were examined by atomic force and scanning electron microscopies, and by energy-dispersive and X-ray photoelectron spectroscopies, respectively. The Hastalex surface was then modified by plasma (3 and 8 W with exposure times up to 240 s), the impact of which on the material surface wettability and morphology was further evaluated. In addition, the material aging was studied at room and elevated temperatures. Significant changes in surface roughness, morphology, and area were detected at the nanometre scale after plasma exposure. An increase in oxygen content due to the plasma exposure was observed both for 3 and 8 W. The plasma treatment had an outstanding effect on the cytocompatibility of Hastalex foil treated at both input powers of 3 and 8 W. The cell number of human MRC-5 fibroblasts on Hastalex foils exposed to plasma increased significantly compared to pristine Hastalex and even to tissue culture polystyrene. The plasma exposure also affected the fibroblasts' cell growth and shape.

**Keywords:** carbon composite; surface modification; pattern; nanostructure; polymer stability; surface chemistry; morphology; cytocompatibility;

*) Corresponding author:

P. Slepička, tel.: +420 220 445 162; E-mail address: petr.slepicka@vscht.cz

## 1. Introduction

In nature, carbon predominantly exists in two allotropes with vastly different physical properties, namely graphite and diamond. However, due to its stability in three hybridizations (sp, sp2, and sp3), carbon can also form numerous other amorphous and crystalline structures [1]. The most promising materials for research are those that have at least one dimension smaller than 100 nm. They are categorized based on their dimensionality as follows: 0D nanostructures including nanoparticles and fullerenes [2], 1D nanofibers and nanotubes comprising carbon nanofibers and carbon nanotubes [3], 2D graphene and nanodots containing graphene, a single layer of carbon atoms, as a prominent example, along with carbon nanodots [4], and 3D structures encompassing carbon foam and carbon aerogel, among others [5]. These carbon nanostructures offer great potential for exploring new materials and their unique properties and applications.

Graphene, which is the fundamental building block of more complex carbon structures, was first successfully prepared in 2004 by graphite exfoliation [6]. Other methods of graphene synthesis include epitaxial growth on silicon carbide [7], chemical vapour deposition (CVD) [8], or dissolution of graphite in organic solvents [9,10]. Nevertheless, a significant limitation of graphene use is its tendency to agglomerate (cluster formation) caused by its high hydrophobicity [11,12]. This property complicates the dispersion of graphene in composite matrices. For practical applications, graphene oxide (GO) [13] and its reduced form (rGO) [14] are often preferred due to their improved dispersibility compared to pristine graphene [15].

GO is prepared by oxidizing graphite [16]; monoatomic layers with oxygen-containing functional groups are formed [17]. The commonly used method for GO preparation is the Hummers method. In this method, flake graphite is treated with sulphuric acid in the presence of sodium nitrate and potassium permanganate [18]. Through this process, graphene enriched in hydroxyl, epoxide, and carboxyl groups is obtained. These functional groups on GO can react with other functional groups and form covalent bonds, thus enabling its incorporation into a matrix during polymerization, for instance [19]. However, during the oxidation process, graphene loses some of its properties [20,21]. As the number of oxygen-containing groups increases, its electrical conductivity decreases. This effect can be partially reversed by chemical [22] or thermal [23] reduction (chemically or thermally reduced graphene oxide, CrGO, and TrGO, respectively).

With the advancing development of carbon nanostructures, they have begun to be considered ideal reinforcements for polymer matrix composites (PMCs) [24,25]. Among them, carbon nanofibers, graphene, and carbon nanotubes (CNTs) have proven to be particularly successful [26-28]. These structures were discovered in 1991, and their tensile strength was measured to be 63 GPa, which is an order of magnitude higher than that of carbon fibers [29]. Polymer composites with CNTs are the most extensively researched composite materials to date, as evidenced by the abundance of scientific publications [30-34]. Due to the strong interfacial interactions of both graphene and CNTs, achieving their dispersion within the matrix is challenging, especially when aiming for higher weight percentages. For graphene, this issue is addressed by using graphene oxide (GO) [17,35,36]. CNTs, on the other hand, are functionalized with carboxylic acid groups to prevent intertwining and bundling [37]. However, some degree of entanglement still occurs. The one-dimensional nature of CNTs leads to inferior material properties when they are freely dispersed, falling short of their theoretical potential. To improve the material properties in a specific direction, methods are being developed to align nanotubes in that direction, such as mechanical stretching [38] or the application of electric [39] or magnetic fields [40]. In the synthesis of carbon-based polymer nanocomposites (PNCs), three main methods are employed depending on the nature of the polymer [41]. These include solvent casting [42], where the polymer is dissolved, and the solvent is subsequently evaporated; melt processing [43], followed by *in situ* polymerization, in which the monomers are polymerized in the presence of the reinforcement, resulting in covalent bonding between the matrix and the reinforcement [42,44]. This is also the case with the GO-based composite material, on which we report in this study.

This new nanocomposite material, traded as Hastalex, is an invention of Prof. A. Seifalian, marketed by NanoRegMed [45]. It is a polymer composite, specifically poly(carbonate-urea) urethane, reinforced with GO. This novel material is being compared to GORE-TEX since both are suitable for medical applications, such as artificial heart valves [46]. Thus far, Hastalex seems to outperform GORE-TEX in this particular application, exhibiting 2.5 times higher tensile strength in one direction and 3.5 times higher tensile strength in the other direction than GORE-TEX. Additionally, it demonstrates better biocompatibility and enhanced resistance to calcification [45,47].

Hastalex has been designed as a biocompatible material for the human body. The last mentioned property is useful for the production of implants that are intended to be present in the body only temporarily, e.g., stents [48]. Hastalex enables gradual tissue healing and subsequent absorption, thus reducing the need for additional implant removal surgeries [49]. Moreover, Hastalex might have potential for further medicinal applications, such as implant manufacturing in orthopaedics, prosthetics production in dentistry, or wound healing support in regenerative medicine [50]. However, importantly, even though Hastalex offers numerous advantages and potential applications, scientific research and clinical studies on this material to assess its efficacy, safety, and long-term success in various settings are still ongoing. At present, only limited research focusing on Hastalex has been done; therefore, the main contribution of this study is to deepen the knowledge about this novel material.

Previously, we have reported some excellent properties of polymer-carbon composite materials, in which the carbon material has been incorporated into the bulk polymer matrix [51-53], grafted onto the polymer surface [54,55], or the carbon layers have been deposited onto the polymer surface and subsequently treated [56,57], which led in all cases in cytocompatibility enhancement and change in cell growth and proliferation as well as cell directionality. In this study, we have focused on a new carbon-polymer composite, which has been studied only for the last two years, and its surface properties induced by several potential surface treatments are still to be discovered. Even a simple surface modification induced by direct current argon plasma with different input power has been applied, the outstanding surface of surface chemistry and moderate changes in surface morphology were observed, while the cytocompatibility of such treated surfaces has been remarkably enhanced.

## 2. Experimental

*Materials and modification*

The studied material was a nanocomposite film (20×20 cm, a thickness of 500 μm) with a polymer matrix and GO reinforcement called Hastalex supplied by Goodfellow. From this film, samples of 1.5×1.5 cm$^2$ were cut and modified by plasma for 10, 40, 80, 120, 160, 200, and 240 s at 3 and 8 W. The material surface exposure to the plasma was done by the BAL-TEC Sputter coater SCD 050 instrument. The exposure was performed using DC cold Ar plasma (gas purity was 99.997%). Process parameters were: Ar flow 0.3 l s$^{-1}$, Ar pressure 10 Pa, electrode area 48 cm$^2$, the inter-electrode distance of 60 mm, chamber volume 1000 cm$^3$. Further, the samples were subjected to thermal stress at 60 and 100 °C using a BINDER drying oven with the fan speed set to level 30/100. The samples were placed onto glass Petri dishes during the thermal exposure.

*Analytical methods*

Gravimetry was used to determine the amount of material removed by the plasma exposure. The prepared samples were weighed five times before and after plasma modification at Mettler Toledo UMX2 automated microbalances with a precision of 0.1 µg.

Next, sample wettability and surface energy were determined by measurement of a contact angle between the substrate and a liquid droplet. For this, SEE System optical device and Advex Instruments software were used. Droplets of water (5 µl) and glycerol (8 µl) were deposited onto the material surface, and photographs of the droplets were taken. The images were then analyzed using a three-point analysis method to obtain the contact angle values. The surface energy values were evaluated by the Owens-Wendt method.

For the study of surface morphology and roughness, the samples were further subjected to atomic force microscopy (AFM) using a Dimension ICON microscope from BrukerCorp. The measurements were conducted in ScanAsyst mode using a silicon tip. The specific probe used was the Scan Asyst Air probe with a spring constant of 0.4 N·m$^{-1}$. Image sizes were captured for an area ranging from 0.3×0.3 µm² to 30×30 µm².

The surface morphology of the substrates was examined and analyzed using a scanning electron microscope (SEM) in a secondary electron mode, specifically, the TESCAN LYRA3 GMU instrument manufactured by TESCAN in the Czech Republic. The SEM analysis was conducted at an accelerating voltage of 5 kV. Additionally, energy-dispersive spectroscopy (EDS) was performed using the same instrument.

Detailed information regarding the elemental composition of the sample surfaces was obtained by energy-dispersive spectroscopy (EDS) using the F-MaxN analyzer and a silicon drift detector (SDD) from Oxford Instruments, UK. The applied acceleration voltage during EDS measurements was 10 kV.

The surface chemistry of the samples was characterized using an X-ray photoelectron spectroscopy (XPS) instrument, specifically the Omicron Nanotechnology ESCAProbeP. Monochromatic X-ray radiation with an energy of 1486.7 eV was used as the excitation source. A surface area of 2×3 mm² was analyzed. The relative abundance of elements was determined from the intensity of peaks in the XPS spectra using CasaXPS software.

To determine the mechanical properties of the Hastalex film, a Discovery DMA 850 instrument from TA Instruments, an American company, was used. This instrument applies a force ranging from 0.1 mN to 18 N, making it suitable for measuring both rigid and soft materials. The device operates in a temperature range of -160 to 600 °C, thus enabling mechanical testing over a wide temperature range.

*Cell Cultivation*

To evaluate the material cytocompatibility, human fetal lung fibroblasts (MRC-5; Merck, USA) were chosen. The cells were regularly passaged using trypsinization and maintained at the exponential phase of growth. Cultivation media consisted of Minimal essential medium (MEM; Merck, USA) supplemented with 2 mM L-glutamine, 1 % (*v/v*) non-essential amino acid solution, and 10 % (*v/v*) fetal bovine serum (all from Thermo Fisher Scientific, USA). The MRC-5 cells were kept in a cell culture incubator at 37 °C, an atmosphere with 5 % $CO_2$ and 95 % humidity.

*Cell Seeding*

The cytocompatibility measurement was done similarly to the refs. [58] and [59], the MRC-5 cells were used from passages 3rd to 5th (from defrosting after delivery). First, the tested materials were sterilized in 70 % (*v/v*) ethanol in water for 1 h, which was followed by rinsing with sterile phosphate-buffered saline (PBS; pH of 7.4). After that, the samples were put into wells of 12-well plates for cell culture (VWR, USA), and hollow PMMA cylinders (Zenit, The Czech Republic) were placed onto them. The MRC-5 cells were then seeded into the individual wells in 1 mL of media containing $1.5 \cdot 10^4$ cells per 1 $cm^2$ in three replicates. The cells were then cultivated at standard conditions for 1, 3, and 6 days. As controls, pristine material and tissue-culture polystyrene (TCPS) were used.

*Cell fixation and staining*

After the endpoint incubation periods of the MRC-5 cells on the tested materials, cell fixation by a crosslinking method occurred similarly as described in ref. [60]. First, the cell cultivation medium was discarded, and, then, the cells were washed two times with PBS, after which fixation with a 4 % (*v/v*) formaldehyde in PBS followed for ca. 20 min. at 23 °C. After this period, the formaldehyde solution was removed, and the cells were washed with PBS and subjected to staining.

*Fluorescent staining*

To visualize inner cell structures as cell cytoskeleton and the nucleus, a phalloidin conjugate and 4′,6-diamidino-2-phenylindole (DAPI) were used, respectively, as described in ref. [61]. Briefly, a solution of 3 µg·$mL^{-1}$ of phalloidin-Atto 488 (AttoTec GmbH, Germany) and 1 µg·$mL^{-1}$ of DAPI (Merck, USA) was added to cells and kept for 20 min. in the dark at 23 °C. After this period, the labeling solution was discarded, and the samples were washed twice with PBS, turned bottom-up since the material was not translucent, and subjected to microscopy.

*Fluorescence microscopy*

The fixed MRC-5 cells were examined using a wide-field inverse fluorescence microscope Olympus IX-81 (Olympus, Japan; xCellence software). The samples were monitored at 100× magnification using a 10× objective with a numerical aperture of 0.30. For fluorescence, a triple, quad filter DAPI/FITC/TRITC (Olympus, Japan) and a high-stability 150-W xenon arc burner (100 % intensity) were utilized. From each sample, at least ten regions of interest were taken, and the images were captured by an EM-CCD camera C9100-02 (Hamamatsu, Japan). Then, the fluorescence images were background-corrected, and the two fluorescence channels (DAPI and FITC) were merged.

*Image processing*

From the images, the number of cells was evaluated based on the nuclei counting using ImageJ 1.54d. First, the images were binarized and thresholded; then the watershed was applied, and the nuclei were counted by "analyze particles". For each sample, ten regions per sample replicate were analyzed. The data were averaged, and the standard deviation was calculated.

## 3. Result and discussion

### 3.1 Surface wettability and aging

First, we prepared a series of samples, which were modified by plasma at 3 and 8 W for 0-240 s, after which their surface wettability and energy were examined using glycerol and water. The contact angles are presented in Fig 2.

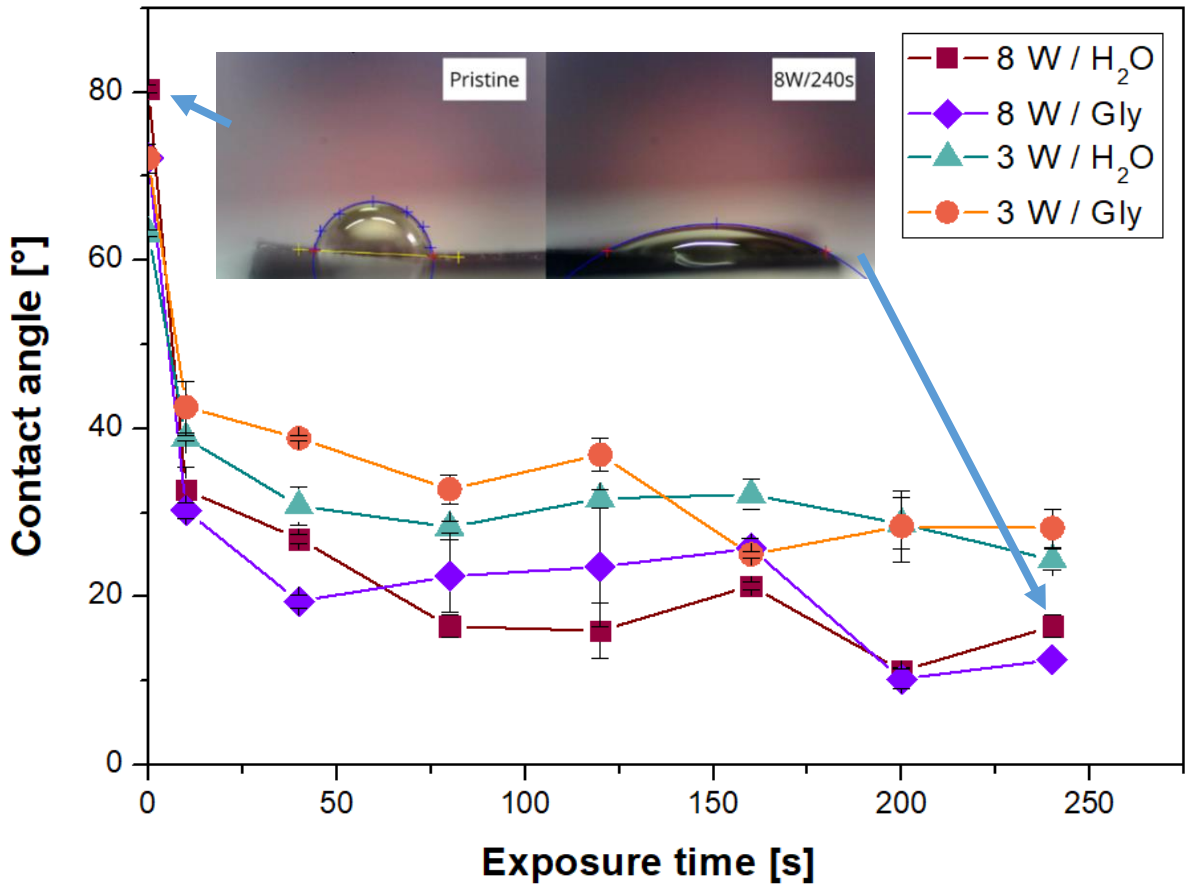

**Figure 1:** Dependence of contact angles between evaluated samples and drops of water or glycerol on the exposure time of plasma at 3 and 8 W.

The graph in Fig. 1 shows that after the initial (plasma) treatment of the material surface, there was a significant change in the contact angle, which became apparent after 10 s. With longer exposure times, the values did not change much; they remained in the range of 15-35 °. Plasma modification at 8 W resulted in a more pronounced change and surfaces with higher wettability compared to the 3 W surfaces. The data in Fig. 2 reveal the dependence of the surface energy of the film on the modification time. As it can be seen, after the plasma treatment with even short exposure time, the surface energy increased from ca. 28 mJ·m$^{-2}$ to values exceeding 60 mJ·cm$^{-2}$.

With longer exposure time, the energy even moderately increased to values slightly above 70 mJ·m$^{-2}$ for an input power of 8 W. It can also be concluded, especially for the higher exposure times, that the input power did not affect the surface energy. With increasing exposure time, this difference disappeared, which was probably due to the surface ablation, which periodically "renews" the surface after a certain period. Therefore, similarly to the trend of wettability data, the surface energy of the films modified with plasma at 8 W was higher than that at 3 W.

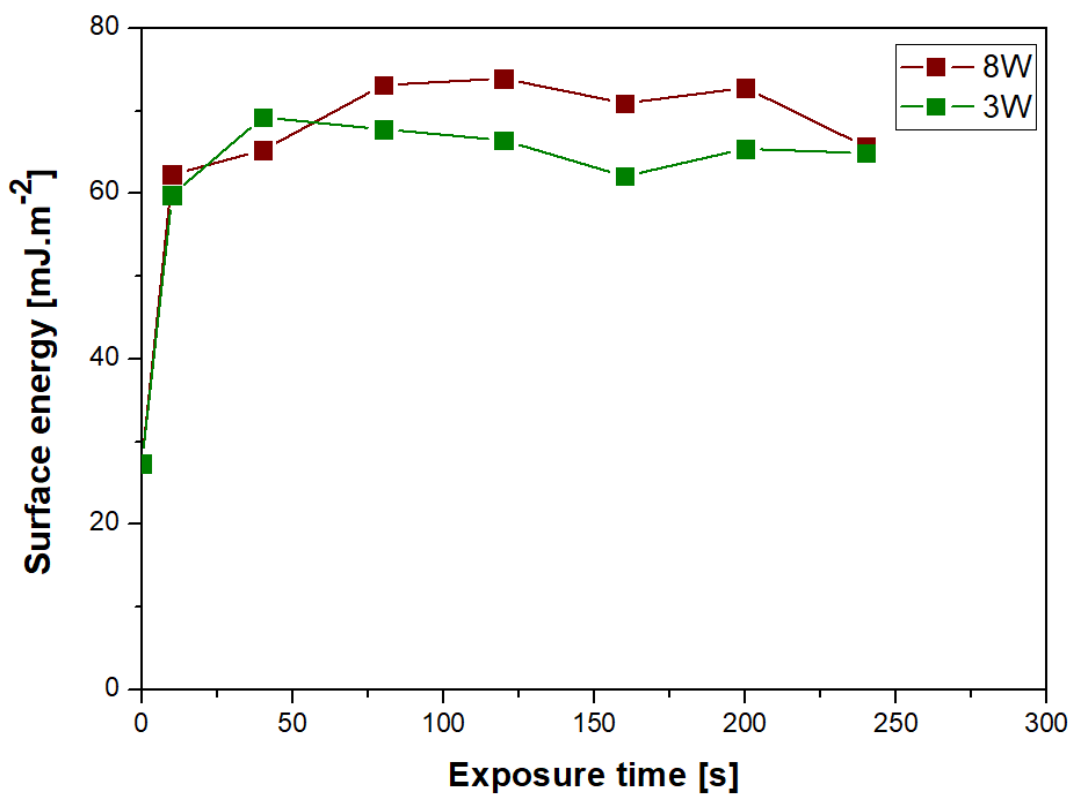

**Figure 2**: Dependence of material surface energy on the exposure time of plasma at 3 and 8 W.

*3.2 Surface morphology*

Next, we have focused on the analysis of surface morphology of pristine Hastalex foil. We have analyzed different surface areas, up to 30 x 30 μm$^2$, and we have chosen to present the data for the four squares of different scales in order to present the morphology of Hastalex from "micro" to "macro" scale. From Fig. 3 A-D, it is apparent that, on the nanoscale, pristine Hastalex is formed by small clusters with extremely low surface roughness (Fig. 3 A); on the basis of our analysis, we also confirm that the plasma treatment induces uniform changes over the treated Hastalex surface. The foil is very flat, and the surface irregularities do not exceed 200 nm in general (on larger scans). The surface roughness ($R_a$) of the film for the examined area of 10×10 μm$^2$ was equal to 22.2 nm (RMS = 29.2 nm).

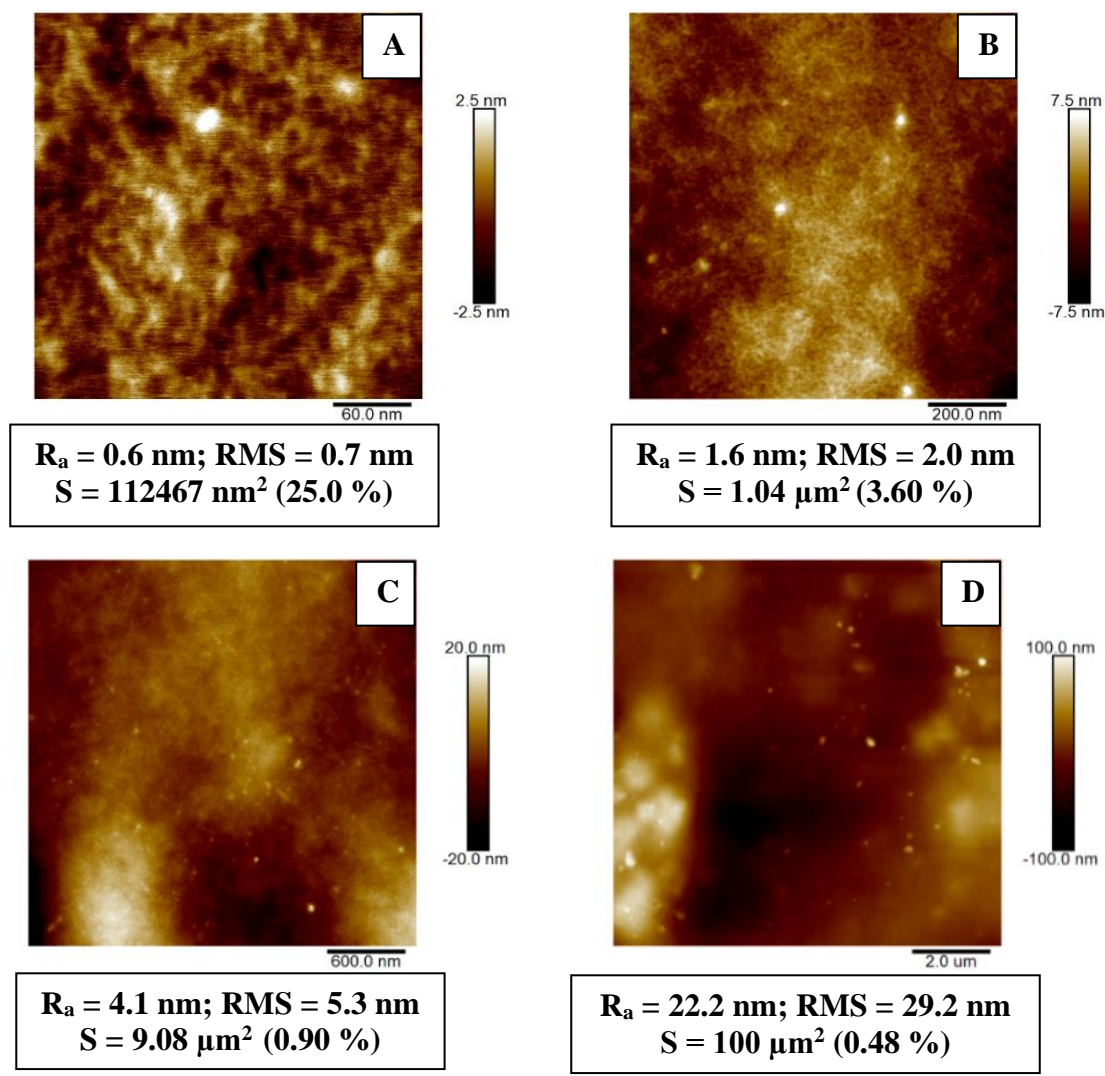

**Figure 3:** AFM images of pristine Hastalex foil (A 300 × 300 nm², B 1 x 1 μm², C 3 x 3 μm², and D 10 × 10 μm²) are shown. $R_a$ indicates arithmetic roughness, and RMS indicates root mean square roughness. S represents an effective surface area with SAD in % (surface area difference).

The AFM scans of the modified material revealed differences in surface morphology between its modifications at 3 and 8 W. In the case of the 3 W/240 s sample, a worm-like structure, characteristic of certain perfluorinated polymers such as perfluoroethylenepropylene (FEP), was already formed on the surface at this applied power and modification time. The presence of these structures was particularly evident in the high-resolution AFM image of 300×300 nm². It can be observed that plasma modification led to an increase in the Hastalex roughness even at a lower power of 3 W. As it is apparent from Fig. 4, after sample exposure to plasma at a power of 8 W, the surface detail exhibited the form of spherical clusters. The higher-power plasma treatment resulted in a more pronounced surface disruption associated with ablation (following section), which led to an increase in surface roughness compared to both the original polymer and the polymer plasma modified at 3 W. Even the lower roughness for higher exposure power was observed, and the effective surface area was dramatically increased. We had in mind that we were still on the very "zoom" on the surface, however, it might still affect cell adhesion to the surface, in combination with altered surface chemistry, which will be discussed further.

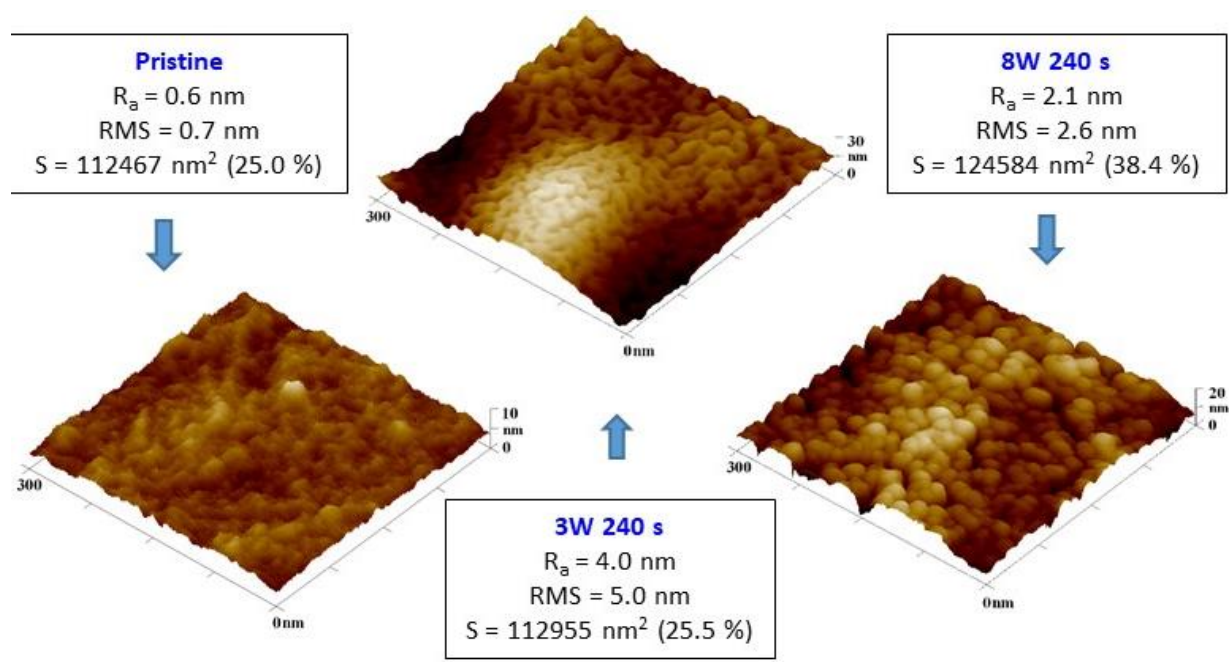

**Figure 4:** AFM 3D images of a Hastalex foil - pristine and modified by plasma at a power of 3 and 8 W for 240 s. A square of 300×300 nm is shown. $R_a$ indicates the arithmetic roughness, and RMS indicates the root mean square roughness, S is the effective surface area.

For further material surface investigation, we employed SEM analysis, the representative images of which are shown in Fig. 5. It is apparent that the samples modified with plasma at both 3 and 8 W exhibited higher roughness and significant morphological changes, which aligns with the results obtained from AFM analysis. We have also determined the surface area up to 30 μm² by AFM technique, but for the sake of clarity, we present results from SEM analysis for larger scanning areas. As it is obvious from SEM and Fig. 5, plasma treatment also induced an increase in surface roughness for Hastalex treated with both 3 and 8 W, but the spheroidal clusters remained on the surface. However, the induced changes in terms of increased surface effective area are also visible on SEM images for plasma-treated samples.

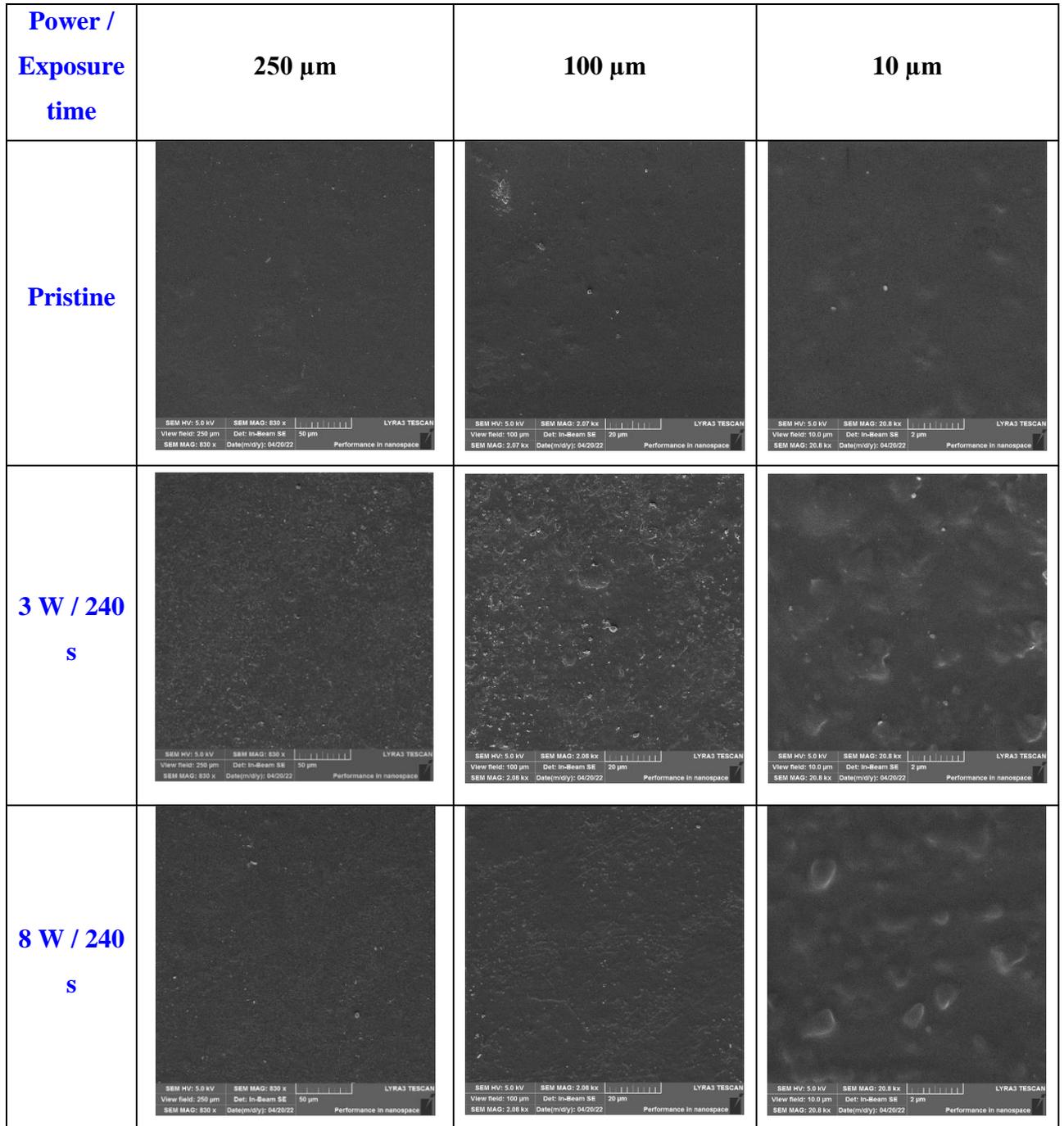

**Figure 5:** SEM images of pristine Hastalex foil and plasma-modified foil with 8 and 3 W with 240 s exposure time. Squares of 10×10 μm$^2$ to 250×250 μm$^2$ are shown.

**3.3 Surface Chemistry**

To determine the elemental composition of the surface layer of both the original and plasma-modified Hastalex samples (up to a few hundred nanometers), energy-dispersive spectroscopy (EDS) was employed. We were particularly interested in the carbon-to-oxygen ratio, which was affected by the material surface exposure to the plasma treatment. We found that the plasma modification led to an increase in the oxygen content, with a more pronounced effect observed at an 8 W power level. Since EDS examines the composition of a material to a greater depth, the difference/impact of plasma between pristine and plasma-treated Hastalex was not as pronounced as it would be if only a few atomic layers were considered (the analysis using XPS will be presented in the following paragraph). The EDS results are shown in Fig. 6.

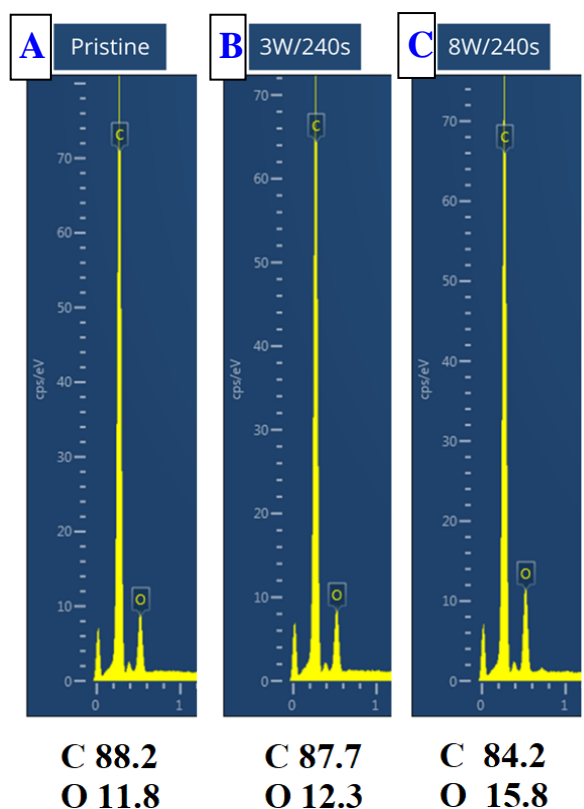

**Figure 6:** EDS spectra of pristine Hastalex foil (A) and Hastalex foil modified with plasma at 3 for 240 s (B) and 8 W for 240 s (C). Carbon and oxygen concentrations in [wt %].

It is evident that the plasma exposure leads to the increase in oxygen surface concentration of Hastalex foil; a more pronounced effect was observed for foils treated with 8 W for 240 s. From Figure 6, it is evident that the argon plasma exposure leads not only to the increase in the atomic concentration of oxygen on the Hastalex foil surface but at the same time to a decrease in the atomic concentration of carbon. It can be concluded that the increase of oxygen-to-carbon ratio confirms the increase of oxygen containing polar functional groups on the Hastalex sample surface. Furthermore, the formation of the oxygen-containing polar functional groups on the Hastalex material surface results in the increase in surface free energy, making the surface hydrophilic at the same time, which explains the decrease in the contact angle after plasma treatment. It is also very important to mention that during plasma surface exposure, ablation took place, as described further in this manuscript. Therefore, partial removal of already modified material was removed from the surface in the form of larger polymeric clusters.

Next, to characterize the composition of the surface-modified layer more accurately, we have applied the XPS analysis for surface chemistry change determination. It was crucial to analyze the surface chemistry of the modified samples as soon as possible after the plasma modification due to rapid changes occurring during surface aging, which may further alter both surface chemistry and/or wettability. The XPS results are presented in Fig. 7. The XPS results indicate a similar phenomenon observed in the contact angle measurements. Differences in the intensities of O1s and C1s spectra peaks were observed. Plasma modification leads to a change in the surface layer. After the modification, the amount of detected O increased slightly, with a more pronounced effect observed at 8 W power. The measurements revealed that plasma treatment enriches the surface layers with oxygen during the modification process.

These outcomes are in good agreement with those observed by the EDS analysis introduced in the previous paragraph. Also, the oxygen concentration was higher than those values determined by EDS since only up to 10 nm of the surface in depth was analyzed, which was more affected by the plasma exposure. Also, the induced difference, the increase in surface oxygen concentration, was significantly higher in the upper layers, which was confirmed by the XPS.

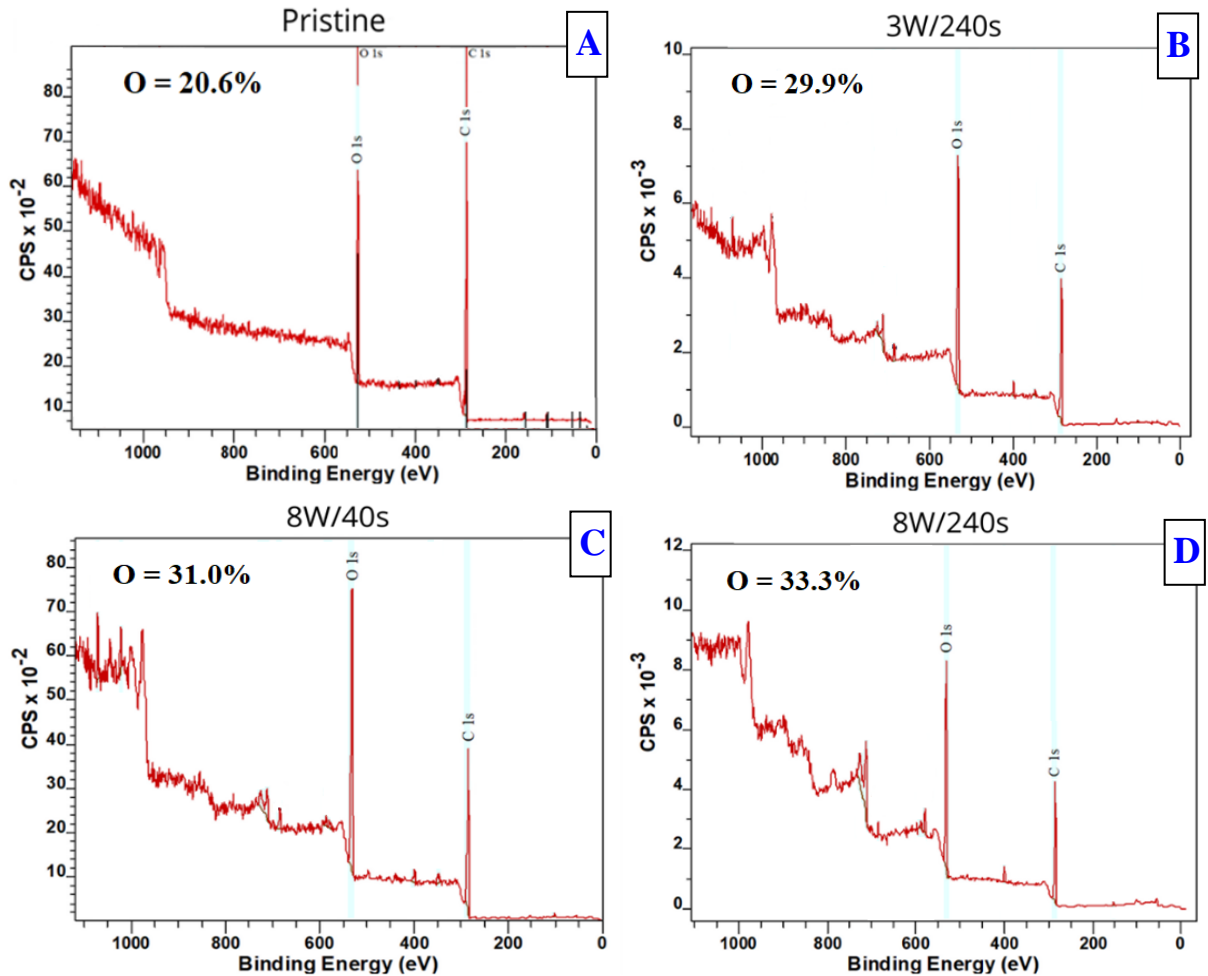

**Fig. 7:** XPS results for unmodified Hastalex foil (A) and Hastalex foil after plasma modification with a power of 3 W for 240 s (B), 8 W for 40 s (C) and 8 W for 240 s (D). Oxygen concentration in at. %.

### 3.3 Mechanical analysis

To assess Hastalex's suitability as a material for heart valves, given its remarkable strength, a dynamic mechanical analysis (DMA) was performed over a temperature range of -100-200 °C at a heating rate of 3 °C per min. The analysis revealed that the glass transition temperature ($T_g$) of the material occurred at -10 °C based on the peak of the tangent delta and -17 °C based on the peak of the loss modulus $E``$. The elastic component of the modulus of elasticity ($E`$) reached values of 103 MPa in the glassy region and 101 MPa in the rubbery region, which are typical for this type of material. No melting temperature or secondary transitions were detected within the temperature range used. The results of the DMA are summarized in Fig. 8.

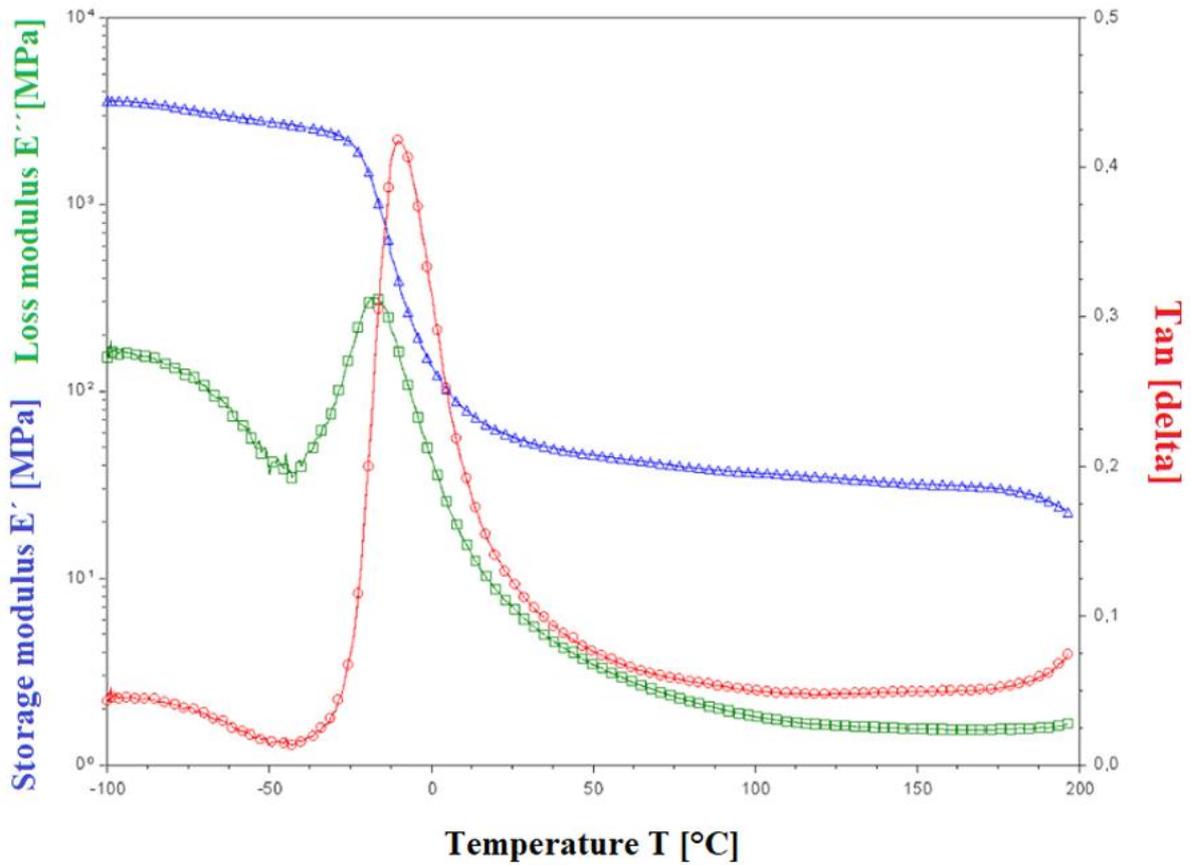

**Figure 8:** DMA analysis of unmodified Hastalex material

### 3.4 Gravimetry

As outlined in the description of the methods, calculating the thickness of the ablated layer was challenging due to the lack of knowledge about the material density during the measurements. To estimate its density, we used the known sample area, accurate mass measurements, and the approximate thickness of the foil, which was provided by the manufacturer with an accuracy of ±10%. We considered this value to be constant, enabling us to obtain a rough idea of the relationship between the thickness of the ablated layer and the exposure time. However, the actual thickness values should be regarded as approximate. This relationship is depicted in Fig. 9.

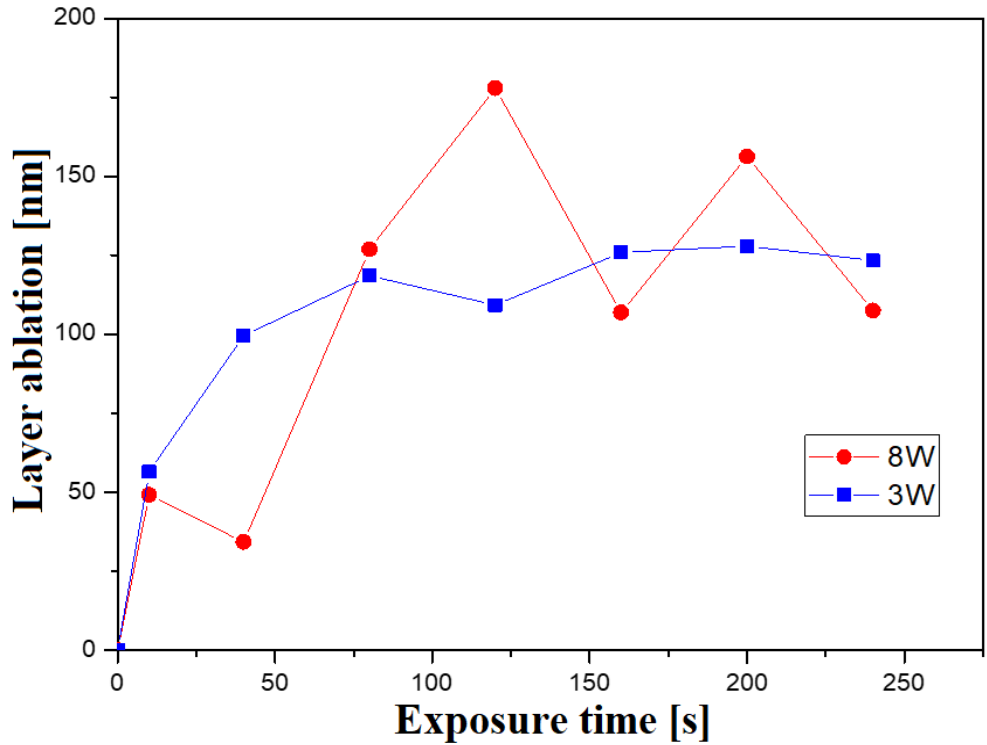

**Figure. 9:** Dependence of foil layer ablation on exposure time for 3 and 8 W input power

Especially after the 3 W modification, it is evident that the initial depletion of the polymeric layer was greater than for longer exposure times. Thus, it is possible to consider the creation of a "passivation coating" on the surface of Hastalex film. As a result of the modification of the plasma, bonds can be cleavage, highly reactive radicals can be formed, and reactions with, e.g., air oxygen can occur. These processes can lead to a change in chemical composition, wettability, morphology, mechanical properties, etc. Some of these parameters are constant over time (they no longer change over time), and some (e.g., wettability) change over time due to the so-called aging of polymers. Elevated temperatures also contribute to this process by acting as accelerators.

**3.5 Heat treatment and aging**

To assess the stability in surface properties, mainly in terms of surface wettability, we heated the Hastalex foil for plasma-treated foil with 8 W for 240 s since these more pronounced changes in both surface chemistry and wettability were detected. After plasma treatment, we analyzed the induced changes in surface wettability. Additionally, such treated foils have been kept in standard conditions (RT) to evaluate the changes in surface contact angle during aging time.

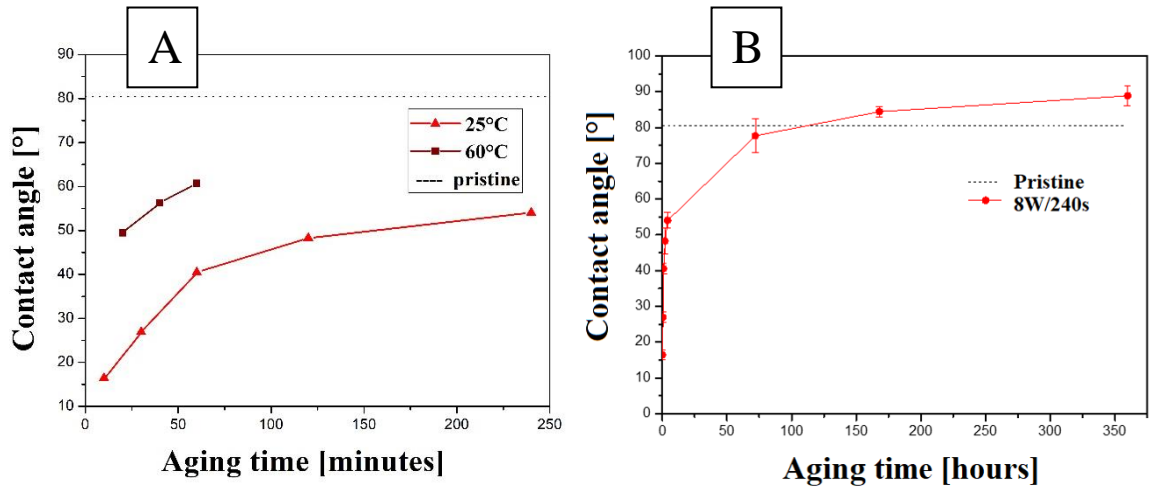

**Figure 10:** Dependence of contact angle on aging time for Hastalex modified 8 W for 240 s and aged under RT (25° C) and 60° C (first 4 hours of aging) (A) and aged under RT in 14 days period (B).

It was found that a temperature of 100 °C was sufficient to fully restore the material surface wettability to the original level of the unmodified material within 1 h, specifically to the value of 81.1 ± 3.7°. Therefore, we reduced the temperature and repeated the measurements at 60 °C, taking a sample every 20 min. and conducting the measurements for 1 h. The results of surface aging at room temperature and elevated temperature are summarized in Fig. 10 A-B. The aging progression during the the initial hours (A) and the 14 days (B) are shown. All these measurements were performed on the most modified film, i.e., with an input power of 8 W for 240 s. It was observed that the wettability returned to the unmodified material value after ca. 125 h continued to decrease.

### 3.6 Cytocompatibility

The interaction of a material with cells can significantly affect its cytocompatibility, which is mostly driven by the material properties, such as roughness, wettability, and composition of functional chemical groups present on the surface. These parameters can be efficiently and economically feasibly modified by plasma and/or laser treatment to obtain a material with tailor-made properties in a controllable manner. This is especially advantageous in the case of inert polymers, such as Hastalex.

Therefore, in this study, human fibroblasts (MRC-5) were chosen as a cell line model to evaluate the cytocompatibility of a Hastalex material before and after its modification with plasma aiming to achieve a material of wettability and surface chemistry which would support cell growth. As a control, tissue culture polystyrene (TCPS) was chosen since it is used as a gold standard for cell cultivation. The material cytocompatibility, cell morphology, and growth of MRC¬5 cells were examined after 1, 3 and 6 days of cultivation, see Figs. 11, 12 and 13, respectively. Cell morphology was monitored based on cytoskeleton staining, and as it is apparent from Fig. 11, after 1 day of cultivation, the MRC-5 cells growing on modified Hastalex were adhered to the material, nicely stretched, and elongated, thus bearing fibroblast-like shapes that corresponded to cells on the TCPS control. In contrast, the cells were mostly round or triangular on the untreated (pristine) Hastalex samples, and only a few were slightly elongated. The shape of these cells was not elongated enough (as in the control) and did not bear the physiologically relevant morphology of fibroblasts. This shape was likely due to the high hydrophobicity of the substrate surface preventing cells from adequately adhering to the material in combination with the "smooth" surface morphology (a surface with very low roughness at the nanometer scale). The change of surface morphology, formation of a surface with higher roughness and globular morphology at the nanometer scale, in combination with the increase in oxygen-containing groups, had a positive effect on cell adhesion on Hastalex polymer, which is apparent from the blue charts in Fig. 12. The adhesion, and the number of cells one day after seeding was similar to control samples on those foils which were treated by plasma. Nevertheless, this positive effect was missing on pristine Hastalex, at which low cell number was determined one day after seeding.

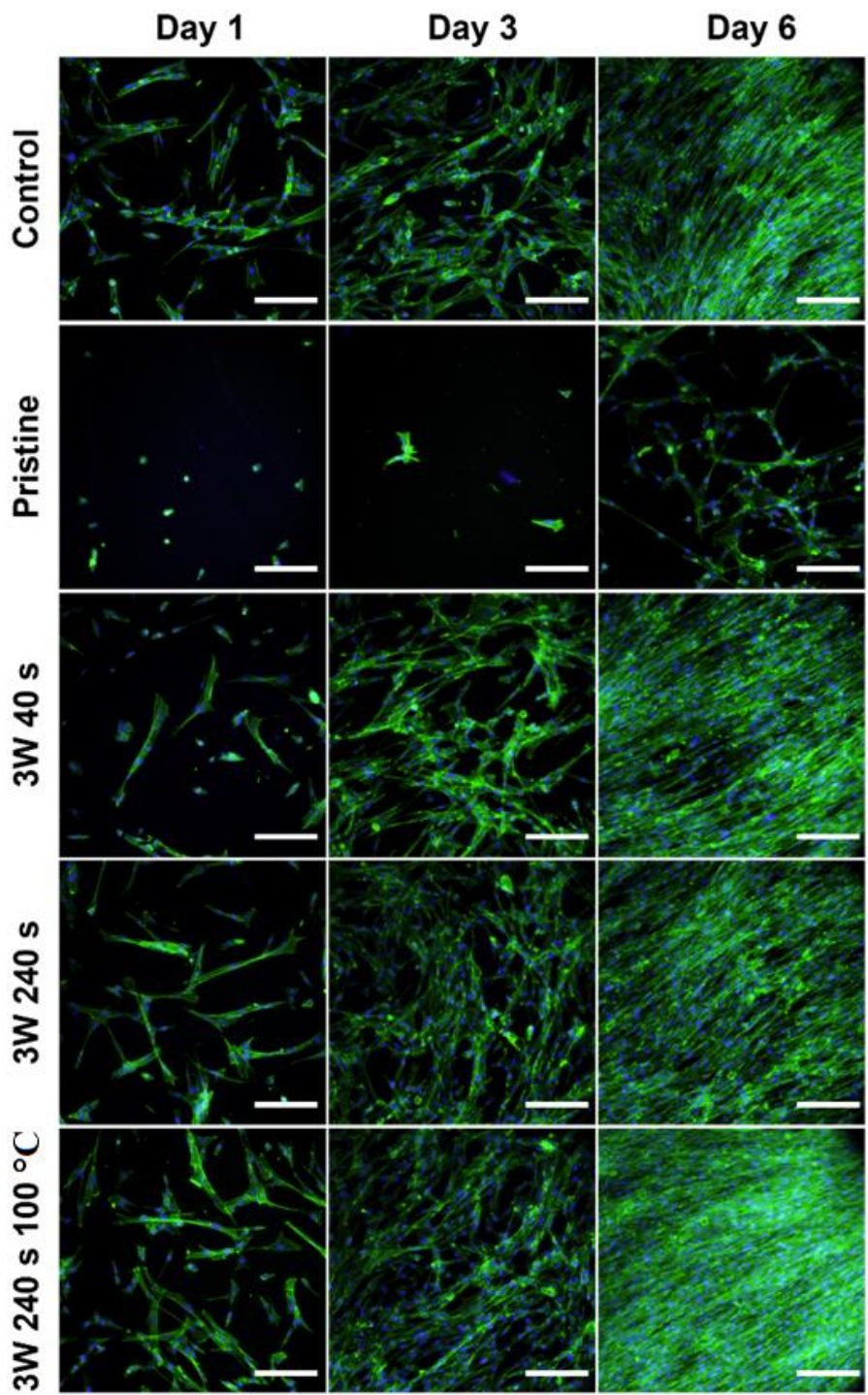

**Figure 11:** A panel of fluorescence microscopy images of human fibroblasts (MRC-5) cultivated for 1, 3, and 6 days on Hastalex and its plasma-modified (3 W) foils. Nontreated Hastalex (pristine) and tissue culture polystyrene (TCPS) served as controls. In green - cell cytoskeleton labeled with phalloidin-Atto488, in blue - cell nuclei stained with DAPI. The scale bar represents 100 μm.

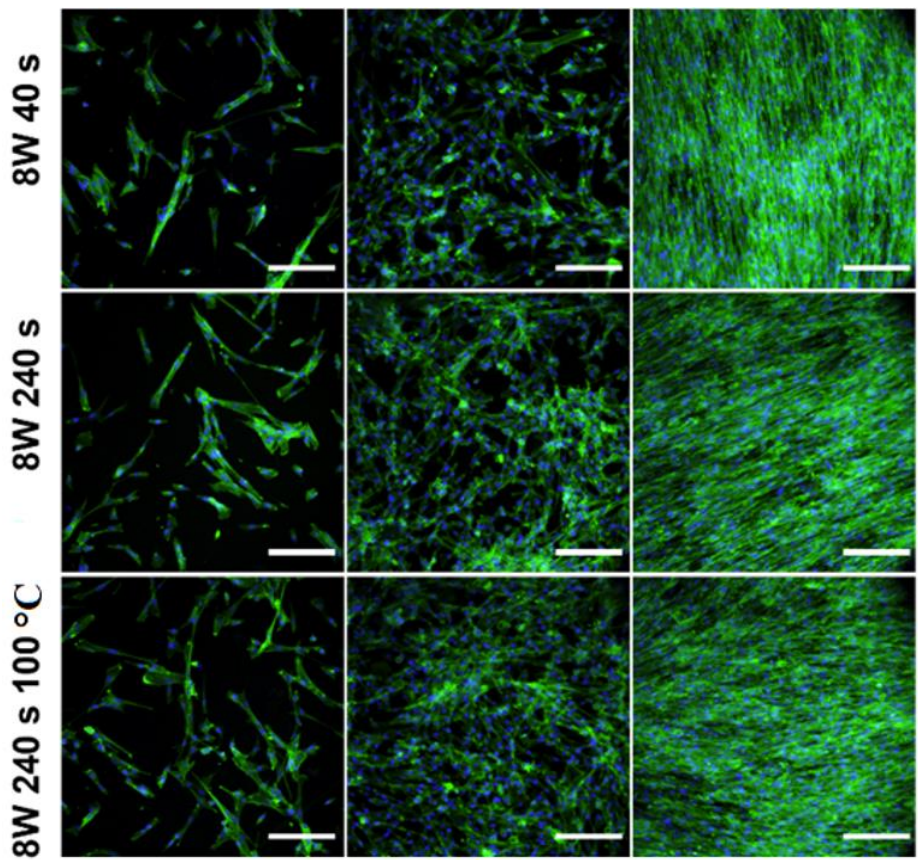

**Figure 12:** A panel of fluorescence microscopy images of human fibroblasts (MRC-5) cultivated for 1, 3, and 6 days on plasma-modified (8 W) Hastalex foils. Nontreated Hastalex (pristine) and tissue culture polystyrene (TCPS) served as controls and are introduced in Fig 12. In green - cell cytoskeleton labeled with phalloidin-Atto488, in blue - cell nuclei stained with DAPI. The scale bar represents 100 μm.

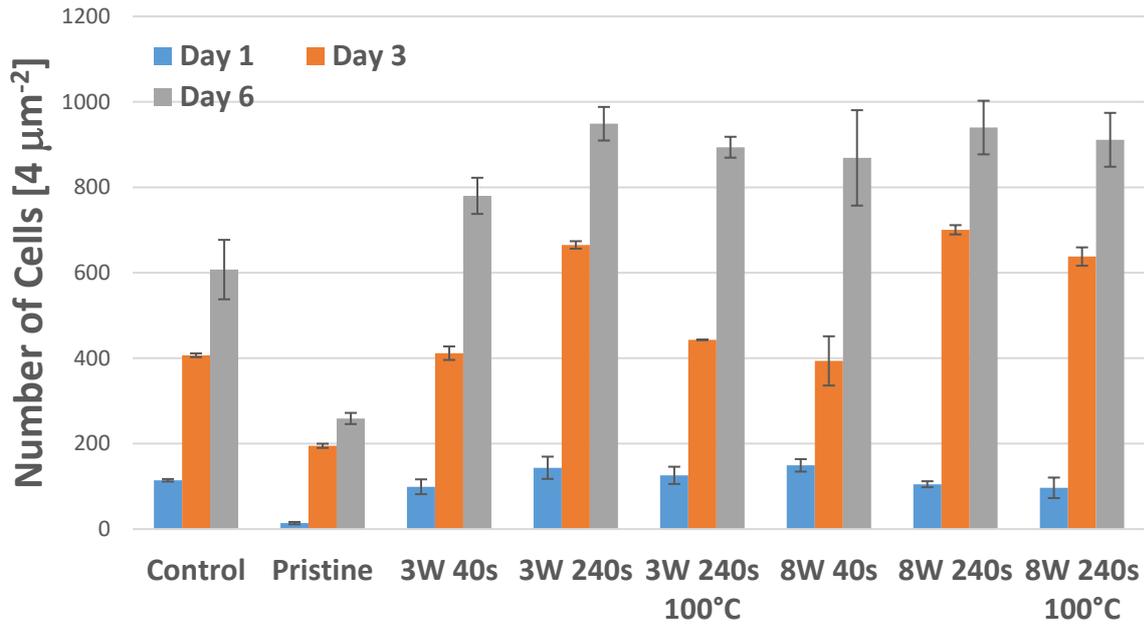

**Figure 13:** The number of human fibroblasts (MRC-5) grown on Hastalex and its plasma-modified variants after 1, 3 and 6 days per 4 μm$^2$. Nontreated Hastalex (pristine) and tissue culture polystyrene (control, TCPS) served as controls. The values are averages of three sample replicates, and the error bars represent standard deviations.

Later, after 3 days of cultivation, the MRC-5 cells growing on modified Hastalex samples still kept their elongated stretched shape and were interconnected with each other. Moreover, dividing cells were also detected on the modified samples. These all were in coherence with cells growing on TCPS control but in big contrast with pristine Hastalex, on which there were only a few cells of irregular shape, mostly triangular and nonelongated. The positive effect of the nanometer scale roughness increase with the oxygen-containing group augmentation is further visible after 3 and 6 days from seeding. Plasma treatment had a positive effect on the number and shape of cells. A higher exposure time of 240 s of argon plasma led to better cell spreading even after three days of inoculation, while samples treated for only 40 s showed lower cytocompatibility. Six days from seeding, the plasma effect on Hastalex cytocompatibility was outstanding, and significantly higher cell numbers were detected even for lower times and powers of plasma exposure compared to a control sample. The cytocompatibility of plasma-treated foils was also maintained after their heat treatment, which is important for different types of sterilization procedures. As for the last cultivation point measured, day 6, the cells formed a confluent 2D layer on all tested samples as well as on control TCPS, except pristine Hastalex, on which there were significantly fewer cells. However, interestingly, the cells were probably already partially adapted to the pristine Hastalex properties since, in addition to a portion of cells growing in clusters of roundish cells, they were mostly of elongated morphology similar to that observed for control TCPS the first day of cultivation, which might be caused by altered surface morphology at the nanometer scale with a lot of adhesion points for filopodia attachment.

## 4. Conclusion

This study is focused on the characterization of a new nanocomposite material based on GO traded as Hastalex. The surface of the unmodified material and the effect of the plasma modification on its properties and morphology were examined. The positive effect of plasma modification on Hastalex surface wettability was demonstrated at both input power but was more pronounced at 8 W. The most significant change in contact angle values occurred within the first 10 s, and the effect of a passivating layer was considered. Next, AFM enabled the detection of material surface roughening after the plasma treatment, and at 3 and 8 W, the formation of worm-like structures and spherical clusters occurred, respectively. To investigate this phenomenon further, we employed SEM, the results of which confirmed that of AFM. XPS, in particular, enabled the composition analysis of the topmost layers of the film. As expected, plasma modification led to an increase in the oxygen content on the material surface. The input power had a greater effect on the oxygen content than the length of the exposure time. Finally, changes in wettability over time following plasma modification were investigated, representing the aging phenomenon. It was observed that wettability decreased to the value of the unmodified material within 125 h at room temperature, and the contact angle continued to increase up to 90° within 14 days. The plasma treatment had an outstanding effect on the cytocompatibility of Hastalex foil for both input power of 3 and 8 W. The cell number on plasma-exposed Hastalex foils increased significantly compared to pristine Hastalex and even to TCPS. The plasma exposure also affected the cell growth and shape; after 6 days from seeding, a dense net of MRC-5 cells was formed.

**Acknowledgments**

This work was supported from the grant of Czech Science Foundation under project 22-04006S, the support of the Ministry of Health of the Czech Republic under grant No. NU20-08-00208 is also gratefully acknowledged. This work was also supported by the Project OP JAK

EXRegmed, No CZ.02.01.01/00/22_008/0004562, of the Ministry of Education, Youth and Sports, which is co-funded by the European Union.

**Data Availability Statement**

The data presented in this study are available at https://doi.org/10.5281/zenodo.10700584.